\documentclass[amsmath,amssym,amsfonts,pre,twocolumn,showpacs]{revtex4}
\usepackage[dvips]{epsfig}

\newcommand{\rv}{{\bf r}}
\newcommand{\nv}{\hat{{\bf n}}}
\newcommand{\bv}{\hat{{\bf b}}}
\newcommand{\ev}{\hat{{\bf e}}}
\newcommand{\xv}{{\bf x}}

\newcommand{\uv}{{\bf u}}

\newcommand{\Rv}{{\bf R}}
\newcommand{\tv}{\hat{{\bf t}}}
\newcommand{\Tv}{\hat{{\bf T}}}
\newcommand{\Nv}{\hat{{\bf N}}}
\newcommand{\Bv}{\hat{{\bf B}}}
\newcommand{\cH}{{\cal H}}
\newcommand{\cT}{{\cal T}}
\newcommand{\cK}{{\cal K}}
\newcommand{\grad}{{\bf \nabla}}

\begin{document}

\title{Braided Bundles and Compact Coils:  The Structure and Thermodynamics of Hexagonally-Packed, Chiral Filament Assemblies}
\author{Gregory M. Grason}
\affiliation{Department of Polymer Science and Engineering, University of Massachusetts, Amherst, MA 01003, USA}

\begin{abstract}
Molecular chirality frustrates the two-dimensional assembly of filamentous molecules, a fact that reflects the generic impossibility of imposing a global twisting of layered materials.  We explore the consequences of this frustration for hexagonally-ordered assemblies of chiral filaments that are {\it finite} in lateral dimension.  Specifically, we employ a continuum-elastic description of cylindrical bundles of filaments, allowing us to consider the most general resistance to and preference for chiral ordering of the assembly.  We explore two distinct mechanisms by which chirality at the molecular scale of the filament frustrates the assembly into aggregates.  In the first, chiral interactions between filaments impart an overall twisting of filaments around the central axis of the bundle.  In the second, we consider filaments that are inherently helical in structure, imparting a writhing geometry to the central axis.  For both mechanisms, we find that a thermodynamically-stable state of dispersed bundles of {\it finite} width appears close to, but below, the point of bulk filament condensation.  The range of thermodynamic stability of dispersed bundles is sensitive only to the elastic cost and preference for chiral filament packing.  The self-limited assembly of chiral filaments has particular implications for a large class of biological molecules -- DNA, filamentous proteins, viruses, bacterial flagella -- which are universally chiral and are observed to form compact bundles under a broad range of conditions.

\end{abstract}
\pacs{87.16.Ka, 61.30.Dk, 87.15.bk}
\date{\today}

\maketitle

\section{Introduction}

From a sufficiently close perspective, understanding the effect of chirality on the organization high-aspect ratio molecules appears to be problem of local geometry.  The lack of mirror symmetry generically implies that rod-like molecules exert a mutual torque, favoring a preferred tilt of neighboring molecules as opposed to a parallel arrangement of molecular axes favored in achiral systems~\cite{harris_kamien_lubensky}.  It is this vantage point that originally compelled Crick to propose the so-called ``coiled-coil" model to explain the structure of $\alpha$-karetin from x-ray diffraction data~\cite{crick}.  But it is well-known that chiral forces, while local in nature, have profound implications for the global arrangement of molecules in ordered systems.  A fascinating example of this global influence occurs in the ``blue phases" of liquid-crystals, wherein mesoscopic chiral order is delicately maintained by a periodic network of topological defects~\cite{wright_mermin}.  A more explicit example of this effect was noted by de Gennes for chiral liquid crystals with smectic order~\cite{degennes_ssc_72}.  Just as the application of magnetic field destroys the superconducting state, the tendency for chiral, or cholesteric, ordering disrupts the ability to form a periodic stack of layers.  At best, chiral order exists in a smectic system in the neighborhood of screw-dislocation grain boundaries, where periodic order has locally been destroyed~\cite{renn_lubensky}.

\begin{figure}
\center \epsfig{file=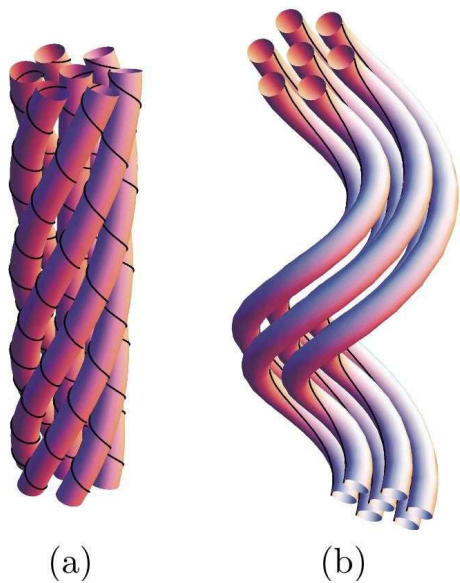, width=2.65in}\caption{Two packing motifs through which molecular structure of constituent filaments imparts chirality on the organization of densely packed bundles.  In (a), intermolecular forces of chiral filaments favor relative tilting of neighboring filaments, leading to an overall twisted bundle.  In (b), the unstressed shape of the filament itself is helical, imparting a writhe to the central axis of the bundle.   }
\label{fig: figure1}
\end{figure}

Beyond the one-dimensional case of smectic materials, chiral-tilt order is more generally incompatible with bulk periodic order.  Indeed, it routinely observed that bulk systems of chiral polymers--such as DNA fragments--expel twist above a critical concentration at which they adopt {\it two-dimensional}, columnar, liquid-crystalline order~\cite{livolant_leforestier}.  For hexagonally-ordered chiral polymers, tilt order that reflects that chirality is only possible in the presence dislocation grain boundaries that locally tear the underlying two-dimensional lattice~\cite{kamien_nelson_95,kamien_nelson_96}.  That columnar phases have this property in common with smectic phases is not a surprise, as the array of two-dimensionally ordered lines can be decomposed into layers, each of which is incompatible with a global twist.

While global chiral ordering is not possible for a {\it bulk} periodic system, the same may not be said for a {\it finite} domain.  This fact is well-appreciated in the context chiral membranes~\cite{prost_helfrich, selinger, ghafouri_bruinsma}, since ribbon of finite width may be twisted into a helicoid without tearing.  Hence, the material is globally ordered in a chiral sense, while at the same time maintaining the inherent layer geometry.   In a recent Letter~\cite{grason_bruinsma_07}, we reported on a similar mechanism in which molecular chirality induces a global twist of a two-dimensional assembly of filaments, provided that the assembly is {\it finite} in diameter.  The chiral ordering of filamentous assemblies has profound implications, with particular consequences for systems of biological filaments, such as DNA or filamentous proteins.  Namely, global twisting of bundles can ultimately limit the stable size bundle aggregates, providing a generic and robust means by which cells may regulate the assembly of a host of biological filaments into fiber bundles.  It is the purpose of this article to present and analyze a general model of filament assembly in which molecular chirality frustrates the ability to form two-dimensionally ordered aggregates, ultimately leading to a state of thermodynamically stable bundle assemblies of finite diameter.

In this article, we explore two geometrically distinct mechanisms through which chirality at the molecular scale frustrates the two-dimensional assembly of filaments.  In the first mechanism, chirality is a colligative effect, favoring a difference in backbone orientation when two filaments are brought into close contact.  Locally, filaments adopt the coiled-coil packing preferred by interlocking molecular screws.  As a filaments are added to a growing bundle, they twist, or braid, helically around the central axis of the bundle (see Figure \ref{fig: figure1} (a)), adopting the so-called ``double-twist" geometry of blue phase liquid crystals~\cite{degennes_prost}.  Such a mechanism was originally proposed to explain the observed twisting of fibrin bundles~\cite{fibrin}, as well as to explain two-dimensionally ordered and twisted structures observed in chromatin of dinoflagellates~\cite{kleman_85}.  In ref. \cite{grason_bruinsma_07}, we established the thermodynamic viability of this mechanism for the first time and explored the sensitivity of the limited bundled growth to exact type and quality of two- and three-dimensional order in the assembly.  

Here, we present a more detailed analysis of the model of twisted, straight bundles.  We find that the optimal degree of bundle twist is a generically non-mononotonic function of bundle radius, with a single maximum whose location is determined only by the mechanical resistance of the bundle to twist.  In the simplest case, that of a hexagonal-columnar, liquid-crystalline bundle, this characteristic size is the bend penetration depth, $\lambda_{3 \perp} \propto (K_3/\mu_\perp)^{1/2}$, where $K_3$ and $\mu_\perp$ are the respective elastic constants describing the resistance to bending and in-plane shear distortions of the hexagonal filament packing.  The predicted non-monotonic dependence optimal twist on bundle radius is critically linked to appearance of an optimal bundle radius, which is in the range of the characteristically mesoscopic length scale, $\lambda_{3 \perp}$.  We find that a dispersed state of finite-sized bundles is predicted to be stable for systems near to, but below, the point of bulk condensation, in which aggregate size is thermodynamically unlimited.

In addition to the case of twisted bundle, here we explore a novel mechanism in which molecular chiral structure obstructs the ability to form two-dimensional filament assemblies.  In this second mechanism, chirality enters through the inherent helical structure of filaments themselves.  This is the well-known geometry of bacterial flagella, whose ``cork-screw" morphology sensitive to solution conditions~\cite{namba} as well as to mechanical stress~\cite{darton_berg_07}.  There is also some evidence to suggest that certain rod-like viruses, like the {\it fd} virus, adopt a weakly supercoiled geometry in presence of certain modifications of the protein coat~\cite{dogic_fraden, tomar_green_day_jacs_07}.  It is already clear from observations of liquid-crystalline mesophases of the {\it salmonella} flagella that helical backbone structure has a dramatic influence on the intermolecular packing of filaments~\cite{barry}.  In the two-dimensional filament packings considered here, the structure of the individual filament imparts an overall writhing geometry to the central axis of the bundle (see Fig. \ref{fig: figure1} (b)).    We consider the case of {\it isometric packings} of filament recently analyzed in detail by Starostin, in which the bundle maintains a perfectly hexagonal arrangement perpendicular the filament backbone~\cite{starostin}.  While these bundles are optimal in terms of constant preferred distance between neighboring filaments, growing a bundle to finite radius requires filaments at the periphery of the to be distorted from their preferred state of writhe.  Thus, we find that the preferred geometry of filaments {\it also} frustrates the growth of hexagonally-packed bundles of helical filaments.

Although this second mechanism is quite distinction from the case where filaments twist around a straight bundle axis -- isometric bundles have strictly {\it no twist} -- the thermodynamic consequences are quite similar.    For solutions of helical filaments sufficiently close to, but below, the point of bulk condensation, bundles of filaments grow to finite diameter.  As the bundle grows, the bundle unwinds, straightening out the helical filaments.  As such, we find that the range of thermodynamic stability of finite-sized helical bundles is sensitive only to the mechanical cost of unbending the filament from its preferred state.  That is, in order to form a bulk aggregate, the net cohesive energy gain per bundled filament, must be larger than the mechanical cost of straightening a helical filament.

This paper is organized as follows.  In Sec. \ref{sec: elasticity} we present a generic, continuum elastic model for hexagonal assemblies of chiral filaments.  In Sec. \ref{sec: twist} we analyze the thermodynamic behavior of hexagonal bundles in the presence of twist-inducing interactions between chiral filaments.  In Sec. \ref{sec: writhe} we analyze a model of helical filaments that form hexagonally-packed, writhing bundles.  We conclude with a brief discussion in Sec. \ref{sec: conclusion}.  

\section{Elasticity of Hexagonally-Packed Filaments}
\label{sec: elasticity}
The structure and mechanical properties of a filament bundle derive from the microscopic interactions between filaments.  While these microscopic interactions may be quite complex, we may understand the inherent cost of long-wavelength distortions of the bundle by considering generic form of the elastic Hamiltonian for two-dimensionally ordered filaments.  These most general description of the elastic properties can be written in terms of $\uv_\perp(\xv)$, which describes the local displacement of a filament from its equilibrium position~\cite{degennes_prost}.  For hexagonally-ordered filaments aligned along the $\hat{z}$ axis, the elastic energy can be constructed from the {\it in-plane} strain tensor,
\begin{equation}
\label{eq: uijperp}
u^{\perp}_{ij} = \frac{1}{2} \big( \partial_i u_{\perp j}+ \partial_j u_{\perp i} - \partial_i \uv_\perp \cdot  \partial_i \uv_\perp - t_i t_j \big) .
\end{equation}
Here, $i$ and $j$ refer only to in-plane directions ($\hat{x}_i \perp \hat{z}$) and $\tv \simeq  \hat{z} + \partial_z \uv _\perp$ is the unit tangent vector describing filament orientation.  The first non-linear term in $u^\perp_{ij}$ is the same non-linear correction appearing for purely two-dimensional solids, while the second non-linear term, familiar from the non-linear elastic theory of smectic liquid crystals, ensures rotational invariance of the elastic energy around in-plane axes~\cite{bruinsma_selinger}.  We will see below that it is necessary to retain the non-linear energy contributions resulting from these ``gauge" terms.  

In the absence of broken translational symmetry along the filament axis, the array responds to elastic deformations as a columnar-hexagonal liquid crystal,
\begin{equation}
\label{eq: Hperp}
\cH_\perp = \frac{1}{2} \int d^3 x \big\{ \lambda_\perp (u^\perp_{kk})^2 + 2 \mu_\perp u^\perp_{ij}u^\perp_{ij} \big\} ,
\end{equation}
where $\lambda_\perp$ and $\mu_\perp$ are the Lam\'e constants describing the resistance to compressive and shear distortions of the hexagonal order perpendicular to the filament axis.  Higher-order derivative contributions take the form the of the Frank free energy for polymer nematics,
\begin{multline}
\label{eq: Ht}
\cH_{\tv} = \frac{1}{2} \int d^3 x \big\{ K_1 (\grad_\perp \cdot \tv)^2+K_2 (\tv \cdot \grad \times \tv)^2 + K_3 |(\tv \cdot \grad) \tv|^2 \\
+K_{24} \grad \cdot [ (\tv \cdot \grad) \tv-\tv(\grad \cdot \tv)] \big\}, 
\end{multline}
where $K_1$, $K_2$ and $K_3$ are the respective elastic constants to splay, twist and bend of the filament backbone.  The total-derivative term on the second line of (\ref{eq: Ht}) is the so-called ``saddle-splay" term, which is often neglected for bulk systems~\cite{degennes_prost}.  For a typical 3D solid, one may ignore the higher-order Frank free energy contributions of eq. (\ref{eq: Ht}), but in the absence broken-translational symmetry along the backbone direction it is necessary to include at least the resistance to bending ($K_3 \neq 0$) in order to stabilize the system to long-wavelength deformations and thermal fluctuations.

Below a critical temperature, a columnar system of filaments will pass to a crystalline state~\cite{grason_08, grelet_prl_08} and adopt the elastic response of a 3D anisotropic solid.  The strain energy of a solid is properly described by the 3D strain tensor~\cite{landau_lifshitz},
\begin{equation}
u_{ij} = \frac{1}{2}(\partial_i u_j + \partial_j u_i - \partial_i \uv \cdot \partial_j \uv) ,
\end{equation}
where $\uv$ here is the three-dimensional displacement and $i$ and $j$ refer to all three spatial coordinates.  The most important distinction between 3D-solid and 2D-columnar response is a resistance to uniform shear along the filament axis.  In addition to the elasticity described by eq. (\ref{eq: Hperp}) a 3D-solid requires the additional elastic terms,
\begin{equation}
\label{eq: Hpara}
\cH_\parallel = \frac{1}{2} \int d^3 x \big\{\lambda_\parallel u_{zz}^2 + 2 \mu_\parallel (u_{xz}^2+u_{yz}^2) \big\} , 
\end{equation}
where, to be clear, this contribution to the elastic response is written in terms of the 3D strain tensor, not $u^\perp_{ij}$.  

The elastic response described by $\cH_\perp + \cH_\parallel$, provides a fully general description of the a 3D hexagonal solid.  By taking either  $\mu_\parallel \neq 0$ or $\mu_\parallel = 0$ we may use this description to model a bundle with either solid or liquid-crystalline order.    Alternatively, we may 
view the composite elastic response in terms the following microscopic model.  Consider a hexagonal array of inhomogeneous filaments where intermolecular forces depend independently on (1) the in-plane separation between neighboring filament backbones and (2) the separation between mass points distributed inhomogeneously along the filament backbone.  The former responses derives, say, from the monopole interaction between charged filaments, while the second results from the combination of multi-pole electrostatic, steric and protein- or ion-mediated forces at work in biofilament systems.  For hexagonally-ordered filaments we must always have $\lambda_\perp\neq 0 $ and $\mu_\perp \neq 0$, while for sufficiently low concentrations of linking agents or sufficiently strong thermal fluctuations $\mu_\parallel$ may be considerably reduced relative to the in-plane moduli.  In the following, we treat these moduli as mesoscopic parameters which may be tuned through underlying microscopic physics.

Finally, we consider the elastic terms that reflect the microscopic chiral structure of biological filaments.  These terms break the manifest chiral symmetry of eqs. (\ref{eq: Hperp}), (\ref{eq: Ht}) and (\ref{eq: Hpara}) and change sign under spatial inversion, $\xv \to -\xv$, but do not change sign under $\tv \to -\tv$.  For filaments possessing hexagonal order the three most relevant chiral elastic terms are,
\begin{multline}
\label{eq: H*}
\cH^* = \int d^3 x \Big\{ \gamma \tv \cdot (\grad \times \tv) +  \gamma' (\tv \cdot \grad) \big[\tv \cdot (\grad \times \uv_\perp) \big] \\ + \gamma'' \tv \cdot \big[  (\tv \cdot \grad)  \grad \times \uv_\perp \big]  \Big\} .
\end{multline}
When $\uv_\perp$ is small, each of these terms is linear in $\uv_\perp$ and second order in derivatives.  The first term is simply the cholesteric twist of the filament backbone typically appearing for chiral nematic liquid crystals~\cite{degennes_prost}.  The second and third terms are rotationally-invariant generalizations of the a chiral symmetry-breaking term appearing in columnar phases, studied by Kamien and Nelson in the context of chiral defect phases~\cite{kamien_nelson_95, kamien_nelson_96}.  To first order in $\uv_\perp$, each of these terms are equivalent and related to the rotation of the hexagonal bond orientation along the column axis, $\partial_z \theta_6$, where $\theta_6$ is the six-fold bond angle of a hexagonal lattice measured with respect to a reference direction in the plane.  Hence, for filament bundles, chirality of the filaments induces ``braiding" of filaments around the bundle axis.  Pursuing these terms to higher order, we note that the final two terms in (\ref{eq: H*}) are different since
\begin{multline}
\grad \times \uv_\perp = 2 \theta_6 \tv + \tv \times (\tv \cdot \grad) \uv_\perp \\ \simeq  (\grad \times \uv_\perp)_z \hat{z} + \hat{z} \times \partial_z \uv_\perp ,
\end{multline}
The second term in  (\ref{eq: H*}) only picks up $\partial_z  (\grad \times \uv_\perp)_z$ while the third has an additional contribution, $\partial_z \uv_\perp  \cdot (\hat{z} \times \partial^2_z \uv_\perp)$.  This higher order contribution is a measure of the writhe of the filament backbone, representing an important class of geometrically distinct chiral deformations.  Specifically, it may be shown that to lowest order in $\uv_\perp$, all three terms in eq. (\ref{eq: H*}) are proportional to the local {\it twist} of filaments around the central axis of the bundle, while $\partial_z \uv_\perp  \cdot (\hat{z} \times \partial^2_z \uv_\perp)$ is proportional (to lowest order) to the  {\it writhe} of the backbone of the bundle.  Twist and writhe terms are more commonly used to describe the geometry of linked curves, such as the two nulceotide strands of a DNA molecule~\cite{fuller_pnas_78}. 
In the Appendix, we show that relationship between the chiral elastic terms of a hexagonally-ordered filament array in eq. (\ref{eq: H*}) can be more formally identified with either the bundle twist, or the local contribution to filament writhe.

In the following sections, we treat these two effects separately.  In Sec. \ref{sec: twist}, we consider the effect of twist-inducing terms, like $\tv \cdot (\grad \times \tv)$, in hexagonal bundles of otherwise straight filaments.  There is an important distinction between the twist- and writhe-inducing terms.  The twisting of filament measures differences in geometry (orientation) between {\it distinct} filaments in the bundle.  Therefore, it is a colligative affect, reflecting the chiral nature filament interactions.  On the other hand, the writhe is really a measure of differences in geometry along a {\it single} filament (see eq. \ref{eq: wr1}).  The latter class of energetic terms would be present {\it even in the absence} of neighboring filaments.  Unlike the twist-inducing terms, the writhe-inducing terms in $\cH^*$ is really a measure of the intrinsically preferred geometry of the filaments themselves.  Therefore, we take an alternative approach to study the assembly of bundles induced to writhe.  In Sec. \ref{sec: writhe} we consider hexagonal assemblies which are induced to writhe by preferred helical structure of the filaments themselves.

\section{Straight Bundles of Twisted Filaments}
\label{sec: twist}

\begin{figure}
\center \epsfig{file=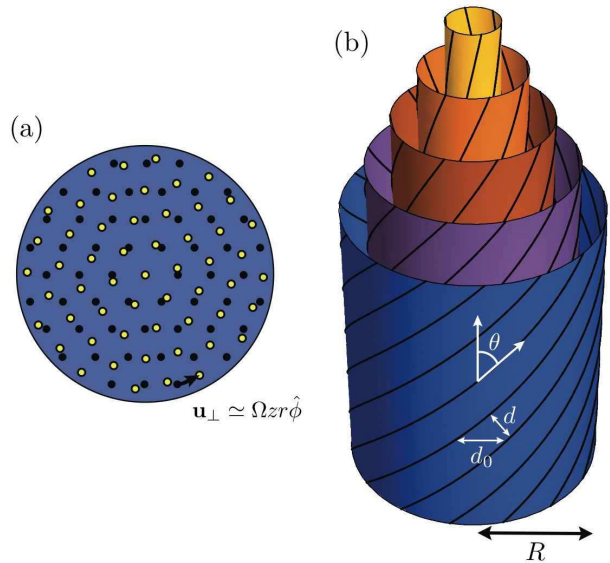, width=3.25in}\caption{(a) shows a cartoon of the torsional deformation of a vertical section of the bundle near $z=0$, demonstrating that the hexagonal packing is largely preserved.  The filled (open) circles show the original (deformed) position of filaments.  (b) depicts cylindrical sections from a twisted bundle at various radii.  Starting from the center and moving outward, according to eq. (\ref{eq: ttwist}), the tilt of filaments relative to the central axis grows approximately linearly.  The tilt angle of outermost filaments is then $\theta = \Omega R$, which leads to a reduction of the spacing between adjacent filaments, $d$, below the preferred spacing, $d_0$.}
\label{fig: twist}
\end{figure}

In this section, we analyze the elastic energy introduced in Sec. \ref{sec: elasticity}, for bundles whose response to the intrinsic chiral stress eq. (\ref{eq: H*}) is to braid around a central axis that is uniformly straight.  This case that was discussed in ref. \cite{grason_bruinsma_07}, in the context of self-limited growth of biofilament bundles.  The torsional deformation of a cylindrical bundle is described by the following in-plain displacement (in cylindrical coordinates)
\begin{equation}
\label{eq: utwist}
\uv_\perp(\xv) = r \big[\cos (\Omega z) -1\big] \hat{r} + r \sin (\Omega z) \hat{\phi} ,
\end{equation}
where $2 \pi / \Omega$ corresponds to the pitch of filaments winding helically around the $\hat{z}$ axis.     From eq. (\ref{eq: uijperp}) it is straightforward to show that this distortion leads to the following in-plain strain,
\begin{equation}
u^\perp_{xx} =-\frac{\Omega^2 y^2}{2}; \ u^\perp_{yy} =-\frac{\Omega^2 x^2}{2}; \ u^\perp_{xy} =\frac{\Omega^2 xy}{2} ,
\end{equation}
hence, the strain-energy grows radially as $(\Omega r)^4$.  Geometrically, this strain energy results from the rotation of the filament tangent, relative to the $\hat{z}$ axis,
\begin{equation}
\label{eq: ttwist}
\tv \simeq \hat{z} + \Omega r \hat{\phi} .
\end{equation}
Figure \ref{fig: twist} shows a bundle whose outer filaments have been rotated to an angle $\theta \simeq \Omega R$ relative the central axis of the bundle.  The distance between filaments along the azimuthal direction is the same as preferred filament separation in the center of the bundle, $d_0$.  Thus, separation of filament {\it perpendicular to the filament axis} at the outer edge of bundles is reduced according to $d^2 \simeq d_0^2 (1-\theta^2/2)$, which indicates that compressive and shear strains of order $\theta^2$ are introduced by the torsional strain.

Depending on the relative magnitude of the in-plane Lam\'e constants, the elastic energy may be lowered by superimposing the displacement of eq. (\ref{eq: utwist}) with an additional dialation in the radial direction according to $\uv_\perp(\xv) \to \uv_\perp(\xv) + (\alpha \Omega^2 r^3/2)  \hat{r}$, where $\alpha$ is variational parameter.   This deformation gives the following contribution to the in-plane elastic energy (correct to $O\big[(\Omega r)^6\big]$ ),
\begin{multline}
\cH_\perp = \frac{1}{2} \int d^3 x \Big\{ \frac{(\Omega r)^4}{4} \big[(\lambda_\perp+\mu_\perp) (1-3 \alpha)^2 \\ +  \mu_\perp(1+2 \alpha)^2 \big] \Big\}. 
\end{multline}
The variational parameter, $\alpha$, is determined by minimizing the elastic energy and hence this parameter depends only on the relative magnitude of the Lam\'e coefficients.  In the limit that the bundle is incompressible in the plane and $\lambda_\perp \gg \mu_\perp$, $\alpha = 1/3$ and for this torsional deformation the elastic due hexagonal order of filaments is $\cH_\perp (\Omega)/V = (25/216) \mu_\perp (\Omega R)^4$, where $V$ is the volume of the cylindrical bundle~\footnote{The numerical coefficient multiplying the in-plane response to twist in ref. \cite{grason_bruinsma_07} was incorrectly reported as $2/27$.}.  Note that no choice of $\alpha$ allows us to relieve the elastic stress introduced by bundle twist.  The elastic strain introduced by a double-twist configuration of lines is a geometric consequence of non-zero saddle splay~\cite{kamien_jphys}, which is incompatible with a constant 6-fold coordinated lattice geometry.

For filament bundles with solid elastic response, the torsional displacement of eq. (\ref{eq: utwist}) also leads to shear contributions to the out-of-plane and divergence-free stress contributions,
\begin{equation}
\sigma_{xz} = \mu_\parallel \Omega y ; \ \sigma_{yz} = - \mu_\parallel \Omega x . 
\end{equation}
The radial force due to these stresses on the boundary of a cylindrical bundle vanishes, as required by mechanical equilibrium.  The linear-elastic response of a solid rod to a torsional deformation is well known, $\cH_\parallel (\Omega)/V = \mu_\parallel (\Omega R)^2 /4$~\cite{landau_lifshitz}.

The Frank-energy contributions can be computed directly from eqs. (\ref{eq: Ht}) and (\ref{eq: utwist}) to show,
\begin{equation}
\frac{\cH_{\tv}(\Omega)}{V} = \Big( K_2 - \frac{K_{24}}{2} \Big)\Omega^2+\frac{K_3 R^2}{4} \Omega^4 .
\end{equation}
Here, we note that the sign of $K_{24}$ is not constrained by symmetry so that for sufficiently large saddle-splay constants ($K_{24} > 2 K_2$), the Frank elastic energy is unstable to twist, even for {\it achiral} systems.  Indeed, model calculations for achiral carbon nanotube ropes suggest that highly adhesive van der Waals interactions drive spontaneous braiding of nanotube ropes~\cite{liang_upmanyu}.  Finally, we summarize the contribution from the chiral-elastic terms in the bundle as $\cH^*(\Omega)/V = -2 \gamma_{{\rm Tw}} \Omega \big(1+O[(\Omega R)^2] \big)$, where 
\begin{equation}
\gamma_{{\rm Tw}} = \gamma + \gamma' +\gamma'' ,
\end{equation}
is the total coupling to the local twisting of filaments around the central axis of the bundle (see Appendix).

The total elastic energy of the twisted bundle is the sum of these components discussed above, $E_{twist} (\Omega) = \cH_\perp(\Omega)+ \cH_\parallel(\Omega)+\cH_{\tv}(\Omega)+\cH^*(\Omega)$.  By defining the following length scales,
\begin{equation}
\label{eq: lambdas}
\lambda_{3 \perp}^2 \equiv  \frac{44 K_3}{25 \mu_\perp}; \ \lambda_{2 \parallel}^2 \equiv \frac{2 (2K_2 -K_{24})}{ \mu_\parallel} 
\end{equation}
and angles
\begin{equation}
\label{eq: thetas}
\theta_{23}^2 \equiv \frac{2 (2K_2 -K_{24})}{K_3} ; \ \theta^2_{\parallel \perp} \equiv \frac{44 \mu_\parallel}{25 \mu_\perp}
\end{equation}
the elastic energy for a straight bundle of chiral filaments takes on the general form,
\begin{equation}
\label{eq: Etwist}
E_{twist}(\theta,R)= \frac{\pi K_3 L}{4} \bigg[ \theta^2 \theta_{23}^2 \Big(1+\frac{R^2}{\lambda_{2 \parallel}^2} \Big) + \theta^4 \Big(1+ \frac{R^2}{\lambda_{3\perp}^2} \Big) - \bar{\gamma} \theta R\bigg]
\end{equation}
where $\theta = \Omega R$ is the twist angle of outermost filaments in the bundle (see Fig. \ref{fig: twist}) and $\bar{\gamma}=8 \gamma_{{\rm Tw}}/K_3$.  For given bundle radius, the equilibrium twist angle derives from the solution to $d E_{twist}/d \theta = 0$.  This solution to this cubic equation predicts the full dependence of bundle twist on the many elastic parameters in model.  Many of these phenomenological elastic parameters may be difficult either to predict from microscopic considerations or to determine independently from macroscopic measurements of filament assemblies.  We focus here on the dependence equilibrium bundle twist, $\theta$, on the radius of the bundle, which may be tested directly by structural observations of bundles under various conditions.  

\begin{figure}
\center \epsfig{file=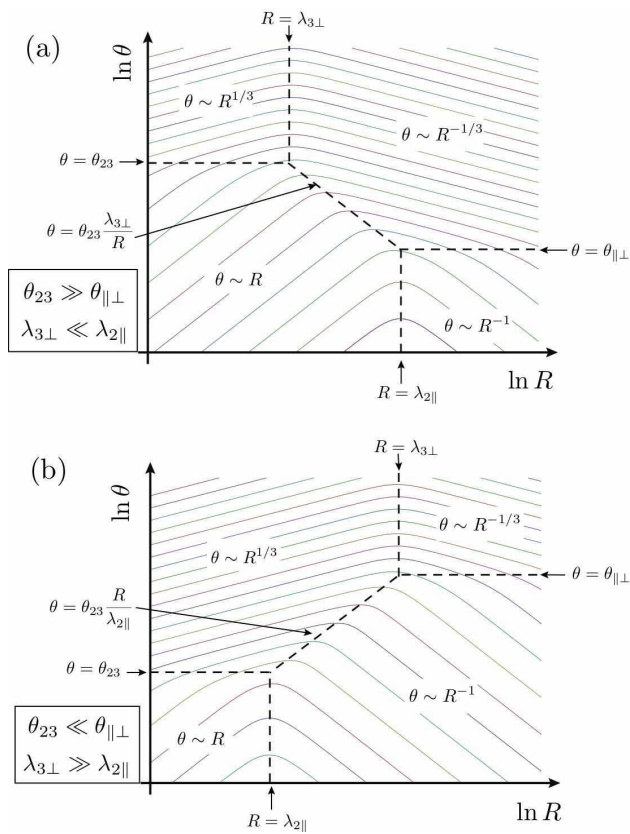, width=3.25in}\caption{Curves predicting the radial-dependence  equilibrium twist angle for fixed elastic constants--according to the minima of eq. (\ref{eq: Etwist})--are shown in (a) and (b).  Each solid curve depicts a different value of the preference from twist $\gamma_{{\rm Tw}}$.  Depending on the relative sizes of elastic parameters $\lambda_{3 \perp}$ and $\lambda_{2 \parallel}$, the variation of $\theta$ with $R$ may be summarized by one of two scenarios, which are shown in (a) and (b) and described by 4 possible scaling regimes.  Note, in particular, that $\theta(R)$ has only a single maximum. }
\label{fig: scaling}
\end{figure}

Despite the many parameters that specify the elastic energy, the mechanical equilibrium allows for a rather limited range of radial dependence of twist.  The chiral term always drives $\theta \neq 0$, while the remaining terms resist torsion.  A simple inspection of eq. (\ref{eq: Etwist}) demonstrates that the form of the dominant elastic restoring force, be it twist, bend or shear stiffness, depend only on the values of $\theta$ and $R$ relative to the angle and length scales defined in eqs. (\ref{eq: lambdas}) and (\ref{eq: thetas}).  The full dependence minima of $E_{twist}$ on $R$ is shown in Figure \ref{fig: scaling}.  

For small bundles, twist generically grows with bundle radius:  for large (small) torsion, bundle twist is restrained by the Frank bend (twist) response.  While for large bundles, equilibrium twist generically decreases with bundle size: at large (small) $\theta$, bundle twist is limited by out-of-plane (in-plane) resistance to shear.  As depicted in Fig. \ref{fig: scaling}, the growth of $\theta$ for small bundles and subsequent decrease for large $R$ only allows for a single maximum in the curve $\theta(R)$.  The non-monotonic dependence of induced twist on bundle size can be directly attributed to the self-limited thermodynamic growth of chiral filament bundles.  The thermodynamic behavior of this model has been summarized in a previous report~\cite{grason_bruinsma_07}.  We therefore present here a brief discussion of two cases:  assembly of columnar-liquid crystalline bundles and assembly of strongly cross-linked, solid bundles.

\subsection{Columnar Liquid-Crystalline Bundles $(\mu_\parallel =0)$}

Above a critical temperature or below a critical density, filaments in a hexagonal bundles are free to slide longitudinally with respect to their neighboring filaments.  This corresponds to the columnar liquid-crystalline state of filament order characterized by a vanishing resistance to shears along the filament axis $(\mu_\parallel =0)$.  In the model discussed above, this corresponds to the case of $\theta_{\parallel \perp} = 0$.  For the sake of simplicity, we also assume in the following that $\theta_{23} = 0$.  The physical basis of this assumption is that the bend elastic constant $K_3$ is determined primarily by the intrinsic stiffness of the filaments themselves, which we expect to be sufficiently large that $K_2/K_3 \ll 1$.  

In this case the elastic energy for a twist bundle takes the form,
\begin{equation}
\label{eq: Ecol}
\frac{E_{twist} (\theta, R)} { \pi L} = \frac{K_3}{4} \theta^4 \Big(1+ \frac{R^2}{\lambda_{3\perp}^2} \Big)-2 \gamma_{{\rm Tw}} \theta R, 
\end{equation}
For this energy it is straightforward to compute the preferred value of bundle twist, $\theta_0$, for a fixed size,
\begin{equation}
\label{eq: the0}
\theta_0(R) = \Big(\frac{ 2 \gamma_{{\rm Tw}} 
}{K_3} \Big)^{1/3} \frac{ R^{1/3} }{(1+R^2/\lambda_{3\perp}^2)^{1/3}} ,
\end{equation}
Note, specifically, the non-monotonic dependence of $\theta_0(R)$.  There is a single maximum at $R = \lambda_{3 \perp}$, and $\theta_0(R) \sim R^{1/3}$ for $R \ll \lambda_{3\perp}$ while $\theta_0(R) \sim R^{-1/3}$ for $R \gg \lambda_{3\perp}$.  Physically, this crossover is due to the fact that bending energy dominates the mechanical resistance to twist for $R <  \lambda_{3\perp}$, while the in-plane shear dominates in the opposite regime.  Using this result in $E_{twist} (\theta_0, R)$ we may compute the radial dependence of the energy gained by the induced twisting of the bundle,  
\begin{equation}
\frac{E_{twist} (\theta_0, R)} { \pi L} =- \frac{3}{2} \Big( \frac{ 2  \gamma_{{\rm Tw}}^4 }{K_3} \Big)^{1/3} \frac{R^{4/3}}{ (1+R^2/\lambda_{3\perp}^2)^{1/3}} .
\end{equation}
Since $E_{twist} (\theta_0, R) \sim - \theta R$, this means that the elastic twist energy gain grows faster than linearly, $-R^{4/3}$, for small bundles  and more slowly than linearly, $-R^{2/3}$, for large bundles.  This result means that the lateral force on the bundle due to elastic forces, $-\partial E_{twist}/\partial R$, vanishes in both the small bundle and large bundle limits.  Under appropriate thermodynamic conditions, this property gives rise to a thermodynamically preferred size, $R_0$, which is finite.

To see this, we consider additional contributions to the free energy of a growing bundle that are due to cohesive forces between neighboring filaments.  In general, we expect that these forces can contribute a net free energy gain per bundled filament, $- \Delta \epsilon L$.  To be clear, $\Delta \epsilon$ is the cohesive energy gain per unit length of {\it parallel filaments} in a hexagonal bundle.  Due to the chiral nature of filament interactions, the twisting of the bundle may further increase the cohesive energy of the bundle, and this mechanism is described by $E_{twist}$.  Additionally, due to the reduced number of adhesive contacts and the edge of the bundle, we may also attribute a positive energy per unit area of the bundle surface, $\Sigma$. 

Combining these twist-independent contributions with $E_{twist}$ we have the total free energy of a bundle,
\begin{multline}
\label{eq: Fcol}
\frac{F(R)}{L} = 2 \pi \Sigma R  - \pi R^2 \rho_0 \Delta \epsilon \\ -  \frac{3 \pi}{2}\Big( \frac{ 2  \gamma_{{\rm Tw}}^4 }{K_3} \Big)^{1/3} \frac{R^{4/3}}{ (1+R^2/\lambda_{3\perp}^2)^{1/3}} ,
\end{multline}
where $\rho_0$ is the number of filaments per unit area of the bundle cross-section.

At equilibrium, aggregates of radius $R$ exist in solution according to a probability $p(R) \propto \exp\{ - F(R)/k_B T\}$~\cite{israelachvili}.  The dominant state of the solution is determined by the global free energy minimum of $F(R)$.  Respective minima at $R\to 0$ and $R \to \infty$ correspond to the states of dispersed single filaments in solution and bulk aggregates of filaments of macroscopic size.  In the absence of any twist-dependent energy gain ($\gamma_{{\rm Tw}} \to 0$), a state of bundles of finite radii dispersed in solution does not occur.    

When $\Delta \epsilon > 0$, then the bundle free energy is unbounded at $R \to \infty$, indicating a state of bulk aggregation (infinite aggregates).  However, when the net cohesive energy for the {\it parallel} state of neighboring filaments is non-positive, two possible states arise.  This is most clear from the case when $\Delta \epsilon =0$ in $F(R)$.  The positive surface energy grows linearly with radius, while the negative chiral-cohesion energy, $E_{twist}$, both vanishes faster than linearly as $R \to 0$ and diverges slower than linearly as $R \to \infty$.  Hence, the surface term generically dominates the bundle free energy in the small and large bundle limits.  If the surface energy is above a critical value, $\Sigma_c$, then the negative cohesive energy gain from bundle twist is insufficient to lower the free energy below the single-filament minimum at $R=0$.  In this case we have a state of dispersed single filaments in solution.  It is straightforward to show that,
\begin{equation}
\label{eq: sigc0}
\Sigma_c (\Delta \epsilon = 0 )= \frac{ 3 K_3 \theta_{max}^{4} }{4 \lambda_{3 \perp}} ,
\end{equation}
where $\theta_{max}=\theta_0(\lambda_{3 \perp})$ is the maximum twist angle of the bundle according to eq. (\ref{eq: the0}),
\begin{equation}
\theta_{max} = \Big( \frac{ \gamma_{{\rm Tw}} \lambda_{3 \perp} }{ K_3} \Big)^{1/3} . 
\end{equation}
When $\Sigma$ is decreased below this critical value, there is a global minimum at some $R_0$ for which $F(R_0)<0$, indicating a thermodynamically stable state of finite-sized bundles.  Just at $\Sigma = \Sigma_c$ the equilibrium size remarkably depends only on the elastic resistance to bundle twist, $R_0 (\Sigma_c) = \lambda_{3\perp}$.  A simple estimate of this length scale may be deduced from the dimensional analysis of filament arrays~\cite{grason_bruinsma_07}.  $K_3$ is largely a measure of the intrinsic bending cost of bending individual filaments, suggesting $K_3 \simeq \rho_0 \ell_p$, where $\ell_p$ is the persistence length of the filaments.  One the other hand the in-plane elastic moduli largely reflect forces between neighboring filaments, governed largely microscopic distances of order $d_0 \simeq \rho_0^{-1/2}$.  Hence, for a columnar array near the limit of thermal stability we expect $\mu_\perp \simeq \rho_0^{-3/2}$.   These estimates suggest that $\lambda_{3 \perp} \simeq (\ell_p d_0)^{1/2}$.  As the persistence length of biological filaments may be on the order of microns, it is clear that the chiral assembly mechanism is quite consistent with the observation of {\it mesoscopic} bundles f-actin which are 10s of filaments in diameter~\cite{wong_prl_07}.  Note also that these scaling arguments suggest an estimate of the critical surface tension in eq. (\ref{eq: sigc0}) in terms of the persistence length:  $\Sigma_c \simeq d_0^{-2} (\ell_p/d_0)^{1/2} \theta_{max}^{4}$ (in units of $k_B T$).  When the surface energy is further lowered below $\Sigma_c$,the preferred bundle size grows, ultimately diverging as $R_0 \sim \Sigma^{-3}$.   

\begin{figure}
\center \epsfig{file=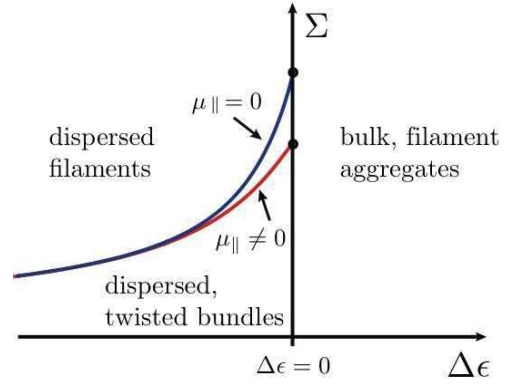, width=2.55in}\caption{The diagram of state for the assembly of chiral filaments into straight bundles of twisted filaments.  When the net free energy gain per unit aggregated filament length, $\Delta \epsilon<0$, filaments are dispersed in solution in one of two states.  Above a critical value of surface energy, $\Sigma_c$, filaments do not aggregate, and for $\Sigma< \Sigma_c$ bundles of finite radius are stable.  The blue curve shows the boundary between these states for columnar, liquid-crystalline bundles and the red curve shows the same boundary for a strongly cross-linked, solid bundle.}
\label{fig: twistphase}
\end{figure}

For the case when $\Delta \epsilon<0$, finite-sized bundles may still be stable below a critical surface energy, albeit with diminished size.  In the limit that $-\Delta \epsilon$ is very large, the bundle size is reduced, ultimately falling of as, $R_0 \sim (-\Delta \epsilon)^{-3/2}$.  This result is obtained by balancing the positive energy penalty, $- \pi R^2 \rho_0 \Delta \epsilon$, with the negative energy gained by twisting very small bundle.  This size scaling also suggests that the critical value of the surface tension falls as $\Delta \epsilon \to - \infty$, since $2 \Sigma_c = R_0 \rho_0 \Delta \epsilon - E_{twist}/R_0 \sim (-\Delta \epsilon)^{-1/2}$.  Therefore, for sufficiently low surface energy and below the point of bulk condensation, this model predicts that a state of dispersed bundles will always be stable.  This thermodynamic behavior is summarized in Figure \ref{fig: twistphase}.

Finally, we note the $\Sigma=0$ behavior of bundles as $\Delta \epsilon $ approached zero from below.   In this limit, the bundle size is much greater than $\lambda_{3\perp}$, with the cohesive penalty, $\pi R^2 \rho |\Delta \epsilon|$, balanced against the chiral cohesion energy in the shear-dominate, large-bundle regime, which grows as $-R^{2/3}$.  Hence, as the point of bulk condensation is approached along the line of vanishing surface energy, the equilibrium bundle diverges as $R_0 \sim |\Delta \epsilon|^{-3/4}$.  In this way, the point $\Sigma=0$ and $\Delta \epsilon=0$ is like a critical point, through which the state of the system may transition continuously from dispersed, finite-sized bundles to bulk aggregates of unlimited size.

\subsection{Solid Filament Bundles $(\mu_\parallel \neq 0)$}

We now briefly consider the effect of a non-zero resistance to shears that slide the filaments with respect to one another along their axes.  Clearly, shear resistance is strong for filamentous protein bundles that are strongly cross-linked by binding proteins.  But we see here that even for the case of a very weak resistance to shear, i.e. $\mu_\parallel/\mu_\perp \ll 1$, the qualitatively different mechanical response to twist produced by $\cH_{\parallel}$ becomes relevant as the size a growing bundle diverges near the point the point $\Sigma=0$ and $\Delta \epsilon=0$.

A non-zero resistance to sliding shears alters the thermodynamics of bundle assembly from the $\mu_\parallel = 0$ case described above in two important ways.  First, the presence of another positive elastic modulus in the model necessarily raises the free energy of a bundle configuration, with a given value of $R$ and $\theta$.  Hence, a non-zero value of $\mu_\parallel$, reduces the range of thermodynamic stability of finite sized bundles.  The most straightforward analysis of this occurs when $\theta_{\parallel \perp}$ is non-zero but sufficiently smaller to consider the effect of out-of-plane shear perturbatively.  For $\mu_\parallel \neq 0$, we consider a more general twist energy,
\begin{equation}
\label{eq: Esol}
\frac{E_{twist} (\theta, R)} { \pi L} = \frac{K_3}{4}\bigg[  \theta^4 \Big(1+ \frac{R^2}{\lambda_{3\perp}^2} \Big) + \theta_{\parallel \perp}^2 \theta^2 \frac{R^2}{\lambda_{3\perp}^2} \bigg]-2 \gamma_{{\rm Tw}} \theta R, 
\end{equation}
where recall from (\ref{eq: thetas}) that $\theta_{\parallel \perp}^2 \propto \mu_\parallel / \mu_\perp$.  If $\theta_{max} \gg \theta$, then the resistance to sliding shear only corrects the predictions of the $\mu_\parallel = 0$ of $R_0$ and $\theta_0$ at $O[( \theta_{\parallel \perp}/\theta_0)^2]$.  Therefore, to leading order the free energy correction per unit length from sliding shear is simply $(K_3/4)  \pi \theta_{\parallel \perp}^2 \theta_0^2 R_0^2$.  From the condition that $F(R_0)=0$ at $\Sigma_c$ we can estimate the leading order reduction of the critical value of surface energy needed to stabilize dispersed bundles,
\begin{equation}
\Delta \Sigma_c = - \frac{K_3 R_0 }{8 \lambda_{3 \perp}^2 } \theta_{\parallel \perp}^2  \theta_0^2 ,
\end{equation}
where $\theta_0$ and $R_0$ are determined according the $\mu_\parallel=0$ model.  At the point of bulk filament condensation the relative size of this correction (from eq. (\ref{eq: sigc0}) is,
\begin{equation}
\Big(\frac{\Delta \Sigma_c }{\Sigma_c} \Big)_{\Delta \epsilon = 0} =  -\frac{\theta_{\parallel \perp}^2 }{6 \theta_{max}^{2} } . 
\end{equation}
The predicted reduction of the thermodynamic stability of dispersed bundles in the presence of non-zero modulus $\mu_\parallel$ in the diagram of state is shown in Figure \ref{fig: twistphase}.

The second principle consequence of a resistance to sliding shears is the reduction in equilibrium size of dispersed bundles.  We consider this effect along the line $\Delta \epsilon$.  As described above for the columnar case, as the bundle surface energy decreases below $\Sigma_c$, the size of the bundle initially decreases as $R_0 \sim \Sigma_{-3}$.  According to eq. (\ref{eq: the0}) this is accompanied by a consequent drop in bundle twist, $\theta_0 \sim R_0^{-1/3}$.  Therefore, even in the limit of vanishingly weak shear response, as the critical point ($\Delta \epsilon = \Sigma=0$) is approached the equilibrium bundle necessarily reaches a point where sliding shear dominates the elastic response, indicated by the fact that $\theta_0$ falls below $\theta_{\parallel \perp}$.  This condition is met for sufficiently large bundles such that, $R_0 \gtrsim \lambda_{3 \perp} (\theta_{max}/ \theta_{\parallel \perp} )^3$.  In this low surface energy limit, the sliding shear dominates the mechanical response to twisted bundle growth.  Hence, equilibrium bundle twist decays with $R$ more rapidly than the $\mu_\parallel =0$ case, $\theta_0 \sim R$.  Given this level of twist, from eq. (\ref{eq: Esol}), the dominant size-dependence of $E_{twist}$ results from the in-plane mechanical resistance which grows as $1/R^2$.  Balancing this energy against the surface energy of the bundle, we find that the growth of the bundle generically crosses over to $R_0 \sim \Sigma^{-1}$ in the $\Sigma \to 0$ limit.  This is a notably weaker divergence than the $R_0 \sim \Sigma^{-3}$ growth of the columnar-hexagonal bundle model.  Figure \ref{fig: rtwist} shows equilibrium dependence of a solid filament bundle radius on the surface energy for a range of mechanical behavior, from $\mu_\parallel \ll \mu_\perp$ to $\mu_\parallel \gg \mu_\perp$.

\begin{figure}
\center \epsfig{file=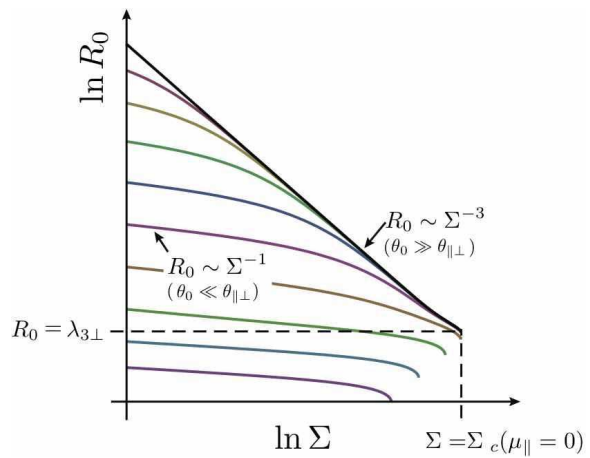, width=3in}\caption{A plot of equilibrium radius of straight bundles of twisted filaments along the line of bulk condensation, where, $\Delta \epsilon = 0$.  For a bundle that is strictly liquid crystalline, $\mu_\parallel=0$, the growth of the bundle is shown by the black curve.  The equilibrium curves spanning the range of weakly solid, $\theta_{\parallel \perp}^2 = 10^{-4}$, to strongly solid, $\theta_{\parallel \perp}^2 = 10^3$, shown in color.  The increased resistance to out-of-plane shear reduces the bundle size, hence, $R_0$ monotonically decreases with  $\theta_{\parallel \perp}^2$.  This figure highlights the $\theta_{\parallel \perp}^2 \neq 0$ crossover for  the divergence of bundle size from columnar response ($R_0 \sim \Sigma^{-3}$) to solid response ($R_0 \sim \Sigma^{-1}$.}
\label{fig:rtwist}
\end{figure}

\section{Writhing Bundles of Helical Filaments}
\label{sec: writhe}

In this section, we explore a separate microscopic model of filament chirality that also leads to the self-limited growth of bundles.  Many biological filaments adopt naturally-helical configuration in their groundstate. The prototypical example of such filaments is the bacterial flagella.  Due to structural confirmations of the protein subunits composing flagella, these filaments are even known to adopt both right- and left-handed helical confirmations~\cite{namba}.  Here, our task is to demonstrate that this type of intrinsic chiral structure also frustrates the assembly of densely-packed filament bundles.

As a microscopic model of intrinsically helical filaments, we consider filaments that are anisotropic in their cross-section and have a tendency to bend around an axis that itself rotates around the filament tangent at a fixed rate.  Following the standard linear elastic description of rod deformations~\cite{landau_lifshitz} we introduce an orthonormal coordinate frame:  $\ev_1 \times \ev_2 =\ev_3 \equiv \tv$.  A filament with a preferred helical configuration can be described by the following elastic energy,
\begin{equation}
\label{eq: Hrod1}
\cH_{{\rm rod}} = \frac{1}{2} \int ds ~\big[ C_{\kappa_1}(\kappa_1 - \kappa_0)^2 + C_{\kappa_1} \kappa_2^2+ C_{\omega}(\omega-\omega_0)^2 \big].
\end{equation}
Here, $\int ds$ denotes an integral over filament arclength and $\kappa_i =   \ev_i \cdot \partial_s \ev_3$ describes bending in the two distinct in-plane directions and $\omega = (\partial_s \ev_1 \times \ev_1) \cdot \ev_3$ describes the rod twist, or rate at which anisotropy of the cross-section rotates around the tangent direction.  $C_1$, $C_2$ and $C_3$ are the bend and torsional moduli that penalize deformations from the ideal, helical state of the filament $\kappa_1 = \kappa_0, \kappa_2=0$ and $\omega = \omega_0$.

When assembled into a hexagonally-coordinated bundle, geometrical constraints make it impossible for filaments to maintain their ideal configuration.  To capture the elastic cost of packing non-ideal filaments we introduce a specific parameterization,
\begin{equation}
\ev_1 = \cos \psi  \nv + \sin \psi   \bv ,
\end{equation}
\begin{equation}
\ev_2 = - \sin \psi   \nv + \cos \psi   \bv ,
\end{equation}
where $\psi$ is a function of filament arc-length, and $\nv$ and $\bv$ are the unit normal and bi-normal of the Frenet frame of the filament backbone~\cite{kamien_rmp_02}.  In these coordinates, the elastic energy of the filament can be computed in terms of $\psi$ and the geometry of the filament backbone,
\begin{multline}
\label{eq: Hrod2}
\cH_{{\rm rod}} = \frac{1}{2} \int ds ~\big[ C_\kappa (\kappa\cos \psi - \kappa_0)^2 + C_\kappa \kappa^2 \sin^2 \psi \\  + C_\omega(\psi'+ \tau -\omega_0)^2 \big],
\end{multline}
where $\kappa$ and $\tau$ are the respective curvature and torsion of the backbone curve, and we consider the simplified case $C_{\kappa_1}=C_{\kappa_2} \equiv C_\kappa$.   For a given backbone geometry, we may compute the induced twist of the filament by solving the Euler-Lagrange equations for $\psi$, which are in the most general case, non-linear and inhomogeneous.  For the purposes of the following analyses, we will consider the two limiting cases.  In the of {\it easy twist}, when $C_\kappa \gg C_\omega $, the filaments lock into the preferred state of bend:  $\psi = 0$.  In the limit opposite limit  of {\it easy bend}, or when $C_\kappa \ll C_\omega $, the groundstate achieves ideal torsion:  $\psi' = \omega_0 - \tau$.  

\begin{figure}
\center \epsfig{file=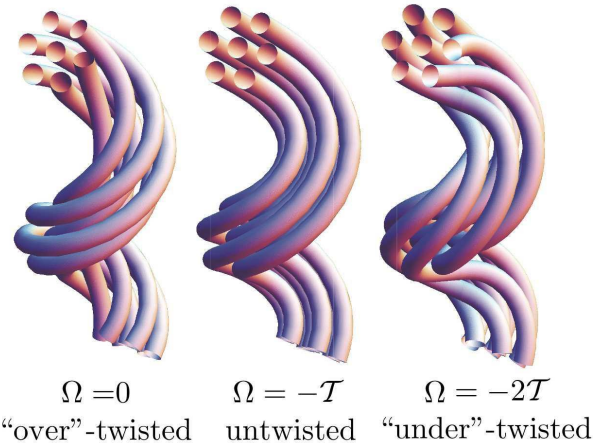, width=3.25in}\caption{Three helical bundles described by eqs. (\ref{eq: bX}) and (\ref{eq: bY}).  An isometric bundle--with a constant separation between all nearest neighbor filaments--can only be obtained if there is no net twist, $\Omega + \cT =0$ (i.e. the coordinate frame compensates for the natural rotation of the Frenet frame).}
\label{fig: isometric}
\end{figure}

In the presence of strong adhesive forces, helical filaments form bundles with a preference for a uniform center-to-center spacing between filament backbones.  Thus, as in the case of straight filaments, in the plane perpendicular to the backbone, the unstressed state of inter-filament forces is hexagonally packed.  The geometry of the bundle can be described in a similar manner to the single-filament geometry.  The central axis of the bundle is described by a central curve, $\Rv_0(S)$, with the associated Frenet frame, $\Tv$, $\Nv$ and $\Bv$.  Note that the upper case vectors refer to the geometry of bundle, while the lower case vectors refer to the geometry of a single filament.  Given the central curve, $\Rv_0$, the positions of the filaments in the bundle correspond to two coordinates, $x$ and $y$, in a coordinate frame that rotates along the bundle trajectory:
\begin{equation}
\label{eq: bX}
\hat{X}= \cos( \Omega S) \Nv + \sin(\Omega S) \Bv,
\end{equation}
and 
\begin{equation}
\label{eq: bY}
\hat{Y} = - \sin(\Omega S) \Nv + \cos( \Omega S) \Bv .
\end{equation}
These coordinates allow for an additional twist of filaments, $\Omega$, relative to the natural rotation of the Frenet frame around the bundle axis.  Using these coordinates the tangents of individual filaments can be derived,
\begin{equation}
\label{eq: twrithe}
\tv = \frac{(\Omega + {\cal T})r \hat{\phi} + (1 - {\cal K} ~ \rv \cdot \Nv) \Tv}{\sqrt{(\Omega + {\cal T})^2 r^2+(1- {\cal K} ~ \rv \cdot \Nv)^2 } } .
\end{equation}
Here, $\rv= x \hat{X}+y \hat{Y}$ is radial separation of filament from the central axis of a bundle and $\hat{\phi}= (-y \hat{X} + x \hat{Y})/r$, is the direction of filament tilt in the plane of hexagonal order.  The respective curvature and torsion of the bundle axis are denoted by $\cK$ and $\cT$.

According to the non-linear contributions to the in-plane strain tensor in eq. (\ref{eq: uijperp}) and eq. (\ref{eq: twrithe}) when $\Omega + \cT \neq 0$, $u^{\perp}_{\phi \phi} \neq 0$, indicating a failure to maintain a constant separation between neighboring filaments in the assembly.  In the case when $\Omega = - \cT$, filaments are collinear; all filaments share the common tangent of the bundle axis, $\tv = \Tv$.  This is condition for an {\it isometric} packing of filaments~\cite{starostin}, in which the interfilament separation is constant in the bundle, throughout the cross-section and along the length.  We should be clear to point out that these structures as distinct from the isometric textures considered by Achard {\it et al.} to model helical ribbon formation in bent-core liquid crystals~\cite{kleman_05} which cannot support a uniform hexagonally geometry in the cross-section.  In the following analysis we focus on this limit of isometric packing favored by strong adhesive forces between neighboring filaments in the bundle.  An isometric bundle is untwisted (i.e. $\tv \cdot (\grad \times \tv) = 0$), distinguishing this mechanism for chiral filament assembly from the one that discussed in Sec. \ref{sec: twist}.

In an isometric packing all filaments share the Frenet frame of the bundle axis, so that differences in filament geometry derive from differences in arc-length element of filaments, $ds$, that of the central bundle axis, $dS$:
\begin{equation}
ds = (1 - {\cal K} ~ \rv \cdot \Nv) dS . 
\end{equation}
From this relation, it is straightforward to compute the relationship between filament geometry and geometry of the writhing bundle,
\begin{equation}
\label{eq: kandt}
\kappa = \frac{ \cK}{1- \cK ~ \rv \cdot \Nv} \ ; \ \tau = \frac{ \cT}{1- \cK ~ \rv \cdot \Nv} .
\end{equation}
These relations reveal the geometric frustration inherent to periodid systems with preferred curvature:  it is not possible to assemble constant-curvature filaments or layers with a uniform separation~\cite{kamien_didonna, kleman_05}.  As bundles grow to larger and larger radii, the curvature and torsion of individual filaments necessarily diverges.  It is this frustration that may lead to a thermodynamic limit to lateral size of a bundle of helical filaments.

\begin{figure*}
\center \epsfig{file=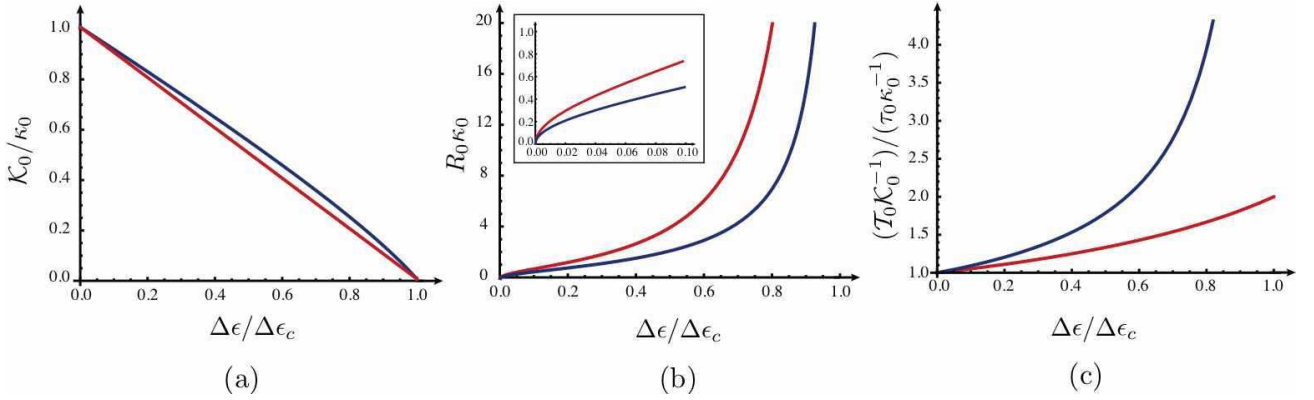, width=6.75in}\caption{Plots of equilibrium structure of bundles of helical filaments for {\it easy twist} model (red curves) and {\it easy bend} model (blue curves).  The dependence of the curvature of the central axis of the bundle, $\cK_0$, on the cohesive energy gain per unit filament length is shown in (a).  The dependence equilibrium size of bundles on cohesive energy is shown in (b), with the inset highlighting the singular growth at the onset of bundling:  $R_0\sim(\Delta \epsilon)^{1/2}$.  The ratio of bundle torsion to bundle curvature, $\cT_0/\cK_0$, is shown in (c).  This ratio corresponds to the ``aspect ratio" of the helical bundle, helical pitch relative to helical radius of bundle axis. } 
\label{fig: writheplots}
\end{figure*}

\subsection{``Easy Twist" Model ($C_\kappa \gg C_\omega$) }

We first consider the case of {\it easy twist}, in which we assume filaments always bend towards the preferred axis, $\ev_1$.   This case arises naturally in the limit that the bending modulus is considerably larger than the torsional modulus of the filaments.  Using the results for the local variation of filament bend and torsion and setting $\psi=0$ in eq. (\ref{eq: Hrod2}), we derive the following elastic cost of an isometric filament packing,~\footnote{When integrating over the circular cross-section of the isometric bundle, care must be taken with volume element as the arclength element $ds$ of individual filaments is not constant.  It is most convenient to choose a coordinate system where $\rv \cdot \Nv = r \cos \phi$, so that $dV = d \phi dr r dS (1-\cK r \cos \phi)$.}
\begin{multline}
\label{eq: easyt}
\frac{E_{{\rm rod}}}{ \pi \rho_0 L}= \frac{ (\cK^2 C_\kappa+ \cT^2 C_\omega)}{\cK^2}\Big[ 1- \sqrt{1-(\cK R)^2 } - \frac{(\cK R)^2}{2} \Big] \\  + C_{\kappa}(\cK-\kappa_0)^2R^2+ C_{\omega}(\cT-\omega_0)^2 R^2 \ \ \ \ {\rm for} \ \psi = 0 .
\end{multline}
Here, $\rho_0$ is number of filaments per unit area in the cross section of the bundle.   The terms in the first line of eq. (\ref{eq: easyb}) account for the singular bending of filaments at the edge of the bundle in limit that $\cK R \to 1$.  This singular dependence of bending energy on $R$ leads to a diverging lateral stress, restraining the lateral growth to $R < \cK^{-1}$.

To analyze the thermodynamics of aggregation we consider the formation of bundles in the presence of cohesive interactions leading to net negative free energy contribution per bundled filament, $-\pi \rho_0 R^2 L \Delta \epsilon$.  Combining the cohesive energy with the elastic cost of packing helical filaments, expanded to $O[R^4]$ we obtain the free energy for growing bundles in the easy twist model,
\begin{multline}
\label{eq: Feasyt}
\frac{F(\cK, R)}{\pi \rho_0 L} = \Big[ \frac{C_\kappa}{2} (\cK - \kappa_0)^2 - \Delta \epsilon \Big] R^2 + \frac{C_\kappa}{8}(\cK R)^4 \\ \ \ \ \ {\rm for} \ \psi = 0 .
\end{multline}
As we are working in the limit that $C_\omega/C_\kappa \to 0$, we have dropped the terms of $O[C_\omega]$.

When $\Delta \epsilon > 0$, filaments aggregate.  However, these aggregates remain finite in radius below a critical value of the cohesive free energy per unit length of bundled filament,
\begin{equation}
\label{eq: depc}
\Delta \epsilon_c \equiv \frac{C_\kappa}{2}  \kappa_0^2 ,
\end{equation}
In the range, $0< \Delta \epsilon < \Delta \epsilon_c$, bundles of finite radii are thermodynamically preferred, while for $\Delta \epsilon > \Delta \epsilon_c$ filaments {\it unwind} form bulk aggregates. 

Minimizing over bundle radii, we find a relation between bundle size, $R_0$, and curvature,
\begin{equation}
R_0^2(\cK) = \left\{\begin{array}{ll} 4 \frac{ [2  \Delta \epsilon-  C_\kappa (\cK - \kappa_0)^2 ]}{C_\kappa  \cK^4}, &   \Delta \epsilon >  \frac{C_\kappa}{2} (\cK - \kappa_0)^2 \\ \\
0, & \Delta \epsilon  <  \frac{C_\kappa}{2} (\cK - \kappa_0)^2 \end{array} \right.
\end{equation}
Using this to rewrite the free energy in terms of bundle curvature only we find for $\Delta \epsilon > 0$,
\begin{equation}
\label{eq: Feasyt2}
\frac{F(\cK, R_0 )}{\pi \rho_0 L} =  \left\{\begin{array}{ll} - 2\frac{ [ C_\kappa (\cK - \kappa_0)^2 - 2 \Delta \epsilon]^2}{ C_\kappa\cK^4 } &  \Delta \epsilon >  \frac{C_\kappa}{2} (\cK - \kappa_0)^2 \\ \\
0 &  \Delta \epsilon  <  \frac{C_\kappa}{2} (\cK - \kappa_0)^2 \end{array} \right.
\end{equation}
Finally, optimizing $F$ in terms of equlibrium bundle curvature, $\cK_0$, we find,
\begin{equation}
\cK_0  = \kappa_0\Big(1-\frac{\Delta \epsilon}{ \Delta \epsilon_c}\Big) .
\end{equation}
From this result we see directly that aggregation leads to the straightening of filaments in the limit that $\Delta \epsilon \to \Delta \epsilon_c$.  In turn, this result yields the following prediction for the equilibrium size of bundles,
\begin{equation}
R_0 =\kappa_0^{-1} \sqrt{ \frac{2  \Delta \epsilon}{ (\Delta \epsilon_c - \Delta \epsilon)^3} } .
\end{equation}
Here, we find that upon aggregation the structure of the bundle evolves {\it continuously} from $\Delta \epsilon =0$ to $\Delta \epsilon =\Delta \epsilon_c$.  As the cohesive strength is increased, the equilibrium value of $\cK$ decreases from the preferred value, $\kappa_0$, to accommodate a greater number of filaments subject to the constraints that $\cK_0 R_0 < 1$.  Note, in particular, the singular growth at the two limits of finite-sized aggregation:  $R_0 \sim (\Delta \epsilon)^{1/2}$ for $ \Delta \epsilon \to 0_+$ and  $R_0 \sim (\Delta  \epsilon_c-\Delta  \epsilon)^{-3/2}$ for $\Delta \epsilon \to \Delta \epsilon_c$.  These results are summarized in Figure \ref{fig: writheplots} (a) and (b).

A further measure of the structural evolution of bundle can be obtained by retaining terms linear in $C_\omega$ in eq. (\ref{eq: Feasyt}) and calculating the evolution of the preferred bundle torsion,
\begin{equation}
\cT_0 = \tau_0\frac{2(\Delta \epsilon_c-\Delta \epsilon  )}{2\Delta \epsilon_c-\Delta \epsilon } .
\end{equation}
This predcits that individuals are untwisted just as they are unbent by stronger cohesive forces.  The ratio $\cT_0/\cK_0$ determines the ``aspect ratio" of the helical bundle, that is, the ratio of the pitch to the radius of the path followed by the central axis of the helical bundle.  The results for $\cT_0$ and $\cK_0$ show explicitly that from the onset of aggregation, when $\cT_0/\cK_0=\tau_0/\kappa_0$, to the point of bulk aggregation, when $\cT_0/\cK_0=2\tau_0/\kappa_0$, the helical shape of the growing bundle continuously stretches along the pitch axis (see Fig. \ref{fig: writheplots} (c)).

\begin{figure}
\center \epsfig{file=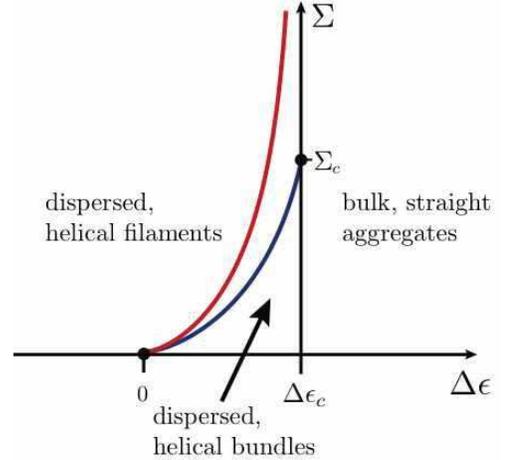, width=2.55in}\caption{The predicted diagram of state for the assembly of helical filaments in solution.  If the net cohesive free-energy gain per bundled filament is below a critical value both dispersed filament and dispersed filaments may be stable, depending on the size of the surface energy of the bundles.  The boundaries between these to states are shown for the {\it easy twist} (red curve) and {\it easy bend} (blue curve) models.  }
\label{fig: writhephase}
\end{figure}

As in the case of straight bundles of chiral filaments, we should also attribute a positive energy cost, $\Sigma$, to the surface of bundle.  Together with the adhesive energy gain per filament, $\Delta \epsilon$, $\Sigma$ determines the thermodynamic stability of finite-radius aggregates of helical filaments.  We estimate a critical value of surface energy, $\Sigma_c$, above which finite-size aggregates are not stable from the condition $2 \pi R_0 L \Sigma_c + F(\cK_0, R_0) = 0$.  This calculation yields the following dependence of $\Sigma_c$ on cohesive energy,
\begin{equation}
\Sigma_c(\Delta \epsilon) = C_\kappa \kappa_0 \rho_0  \frac{ (\Delta \epsilon / \Delta \epsilon_c)^{3/2}}{4  \sqrt{1-(\Delta \epsilon / \Delta \epsilon_c)} } .
\end{equation}
Because $\Sigma_c$ diverges as $\Delta \epsilon \to \Delta \epsilon_c$, the easy twist model predicts that as the system approaches the point of bulk aggregation from the state of dispersed single filaments, it necessarily passes state of dispersed bundles of finite diameter.  The predicted diagram of state is shown in Figure \ref{fig: writhephase}.

\subsection{``Easy Bend" Model ($C_\kappa \ll C_\omega$) }

We now consider the opposite case of {\it easy bending}, in which filaments lock into the the preferred state of twist despite the constraints imposed by isometric filament packing in the bundle.  From eq. (\ref{eq: Hrod2}), this is accomplished when $\partial_s \psi= \omega_0 - \tau$.  Using the result of eq. (\ref{eq: kandt}) for the variation in filament torsion and setting $\cT = \omega_0$, yields the following filament rotation $\psi = -\kappa r \sin(\phi + \omega_0 S)$.  
\begin{multline}
\label{eq: easyb}
\frac{E_{{\rm rod}}}{ \pi \rho_0 L}= C_\kappa \Big[ 1- \sqrt{1-(\cK R)^2 } - \frac{(\cK R)^2}{2} \Big] \\  - 2 C_\kappa \frac{\kappa_0}{\cK}  \Big[ \cK R  ~ J_1 (|\cK| R) - \frac{(\cK R)^2}{2} \Big]  + \frac{C_{\kappa}}{2}(\cK-\kappa_0)^2R^2  \\  \ \ \ \ {\rm for} \ \psi' = \omega_0 - \tau .
\end{multline}
Note that in comparison to the $C_ \omega=0$ ,easy-twist model in eq. (\ref{eq: easyt}), this result only differs by the term in square brackets on the second line.  This is a signature of a more rapid increase of the excess bending induced by the bundle geometry in this limit

Including the cohesive energy gain per unit length of filament and expanding this result in the limit that $\cK R \ll 1$ as before we obtain,
\begin{multline}
\frac{F(\cK, R)}{\pi \rho_0 L} = \Big[ \frac{C_\kappa}{2} (\cK - \kappa_0)^2 - \Delta \epsilon \Big] R^2 + \frac{C_\kappa}{8}(\cK^4+ \kappa_0 |\cK|^3) R^4 \\ \ \ \ \ {\rm for} \ \psi' = \omega_0 - \tau .
\end{multline}
Again, notice that the correction due to imperfect filament packing in the easy bend grows as $|\cK|^3$ rather than the weaker $\cK^4$ response of the easy twist model.  Proceeding as before we minimize the free energy over bundle radii to find,
\begin{equation}
R_0^2 = \left\{\begin{array}{ll} 4 \frac{2  \Delta \epsilon-  C_\kappa (\cK - \kappa_0)^2}{C_\kappa(\cK^4+ \kappa_0 \cK^3)} & 2  \Delta \epsilon >  C_\kappa (\cK - \kappa_0)^2 \\ \\
0 & 2  \Delta \epsilon  <  C_\kappa (\cK - \kappa_0)^2 \end{array} \right.
\end{equation}
and,
\begin{equation}
\label{eq: Feasyb2}
\frac{F(\cK, R_0 )}{\pi \rho_0 L} =  \left\{\begin{array}{ll} - 2\frac{ [ C_\kappa (\cK - \kappa_0)^2 - 2 \Delta \epsilon]^2}{ C_\kappa(\cK^4+ \kappa_0 \cK^3) } & 2  \Delta \epsilon >  C_\kappa (\cK - \kappa_0)^2 \\ \\
0 & 2  \Delta \epsilon  <  C_\kappa (\cK - \kappa_0)^2 \end{array} \right.
\end{equation}
As before there is an upper limit to cohesive energy at which finite-sized bundles will form that is determined only by the energy cost per unit length of unbending the helical filaments:  $\Delta \epsilon_c = C_\kappa \kappa_0^2/2$.  Unlike the easy twist case, it is not possible solve for preferred bend analytically.  Nevertheless, it is possible to analyze bundle properties near the respective points of bundle formation ($\Delta \epsilon \to 0$) and of bulk aggregation ($\Delta \epsilon \to \Delta \epsilon_c$).  

It is straightforward to show that in the limit of $\Delta \epsilon \to 0_+$, $\cK_0 \simeq \kappa_0$ minimizes $F(\cK,R_0)$.  Hence, the equilibrium bundle size grows continuously from $\Delta \epsilon = 0$ as $R_0 \sim \Delta \epsilon^{1/2}$, with the same exponent as the easy-twist model.   Again, as cohesive strength increases and the size of the bundle grows, the bundle curvature is diminished, so that in the opposite limit, near the point of bulk condensation, $\cK_0 \to 0$ linearly as $(\Delta \epsilon_c - \Delta \epsilon )\to 0$.  Thus, as before when $\Delta \epsilon \to \Delta \epsilon_c$, the equilibrium size of bundles diverges, albeit with a smaller exponent than in the easy twist case:  $R_0 \sim (\Delta \epsilon_c - \Delta \epsilon)^{-1}$.  Unlike the easy twist case, however, bundle torsion maintains the preferred values, $\cT_0 = \omega_0$, for condition all cohesive strengths.   Therefore, the aspect ratio of the helical bundle (pitch/radius) as characterized by $\cT_0/\cK_0$, divergences as point of bulk condensation is reached, indicating the conformation of the bundle to be highly extended.  This behavior is summarized in Figure \ref{fig: writheplots} (a)-(c).   

Finally, we compute on estimate for the maximum surface energy, $\Sigma_c$, for which finite-sized bundles are thermodynamically preferred over single helical filaments.  The predicted diagram of state is shown in Figure \ref{fig: writhephase}.  In contrast to the easy twist model, we find a maximum critical surface tension at the point of bulk condensation determined only by the preferred structure and bending elastic of a single filament:
\begin{equation}
\Sigma_c (\Delta \epsilon_c) =\frac{C_\kappa \kappa_0 \rho_0}{3 \sqrt{3} }  .
\end{equation}
Interestingly, in either the easy twist and easy bend limits, the assembly properties of the helical, isometric bundles is sensitive only the elastic cost of deforming filaments from their preferred state of bend, and not to $C_\omega$ or $\tau_0$.  From Fig. \ref{fig: writhephase}, however, we see that the resistance to filament twist has important consequences for equilibrium assembly of helical filaments.  Tuning the ratio $C_\omega/ C_\kappa$ from 0 (easy twist) to $\infty$ (easy bend) leads to a drastic reduction in the range of thermodynamic stability of dispersed bundles of finite diameter.

\section{Conclusion}
\label{sec: conclusion}

In this study, we have explored two geometrically distinct mechanisms by which chiral structure frustrates the two-dimensional assembly of filaments.  In the first mechanism, forces between chiral molecules favor a preferred relative twist of the orientation of neighboring molecules.  While such a twist would be incompatible with a bulk system of hexagonally-ordered filaments, a globally twisted structure is compatible to two-dimensional packing in bundles of finite diameter.  The competition between the effect of chiral interactions and the various mechanical costs of distorting the hexagonal filament packing give rise size-dependence twisting of filaments around the central axis of the bundle.  The size-dependent twist of the bundle, in turn, leads to a thermodynamically preferred, finite lateral size of the bundle, provided that adhesive forces between filaments are sufficiently weak.   

In the second mechanism, we consider the assembly of filaments whose preferred state is one of helical structure.  The complexation of helical filaments induces a natural writhe to backbone of a growing bundle.  Unlike the case of where interfilament forces induce a relative twist of molecular orientation, this mechanism allows filaments to form without an elastic cost for distorting the hexagonal packing in the plane perpendicular to the filament tangents.  The isometric packings preserve the nearest-neighbor separation between all filaments in the bundle.  As a consequence of the preferred writhing the geometry of the bundle, in order for a bundle to grow filaments must be distorted, bent and twisted, from their preferred geometry, ultimately leading to a singular elastic cost as $\cK R \to 1$.  This elastic cost of distorting filaments provides a thermodynamic limitation to the size growing bundles provided the adhesive energy gain per bundle filament is less than the mechanical cost required to unbend a filament, $\Delta \epsilon_c$.  

In many ways, eukaryotic organisms make their living through the constant assembly and disassembly of filamentous molecules.  Functions as diverse as cell division and locomotion are accomplished by means of relatively small class of filamentous proteins:  f-actin, microtubules and intermediate filaments~\cite{alberts}.  {\it In vitro} studies of the assembly behavior of filamentous proteins have contributed greatly to the understanding of physical mechanism by which cells regulate cytoskeletal assembly and function~\cite{theriot}.  In particular, it has been widely noted that when forces between filaments in solution are sufficiently attractive, dense bundles of aligned filaments form, with a limited diameter~\cite{tang_pchem_96, pelletier_prl_03, wong_prl_07, hosek, cebers, purdy, bausch}.  This is most surprising in view of apparent absence of long range forces between filaments in solution.   According to the thermodynamic description of such an assembly process, the classical nucleation model, there can be no equilibrium limitation to the growth in the presence of a net free energy gain per aggregated filament.  To explain this apparent contradiction, a number of theoretical mechanisms for the self-limited growth have emerged, focusing variously on the specialized nature of forces between filaments condensed in the presence of multi-valent counterions~\cite{ha_liu, henle} as well as the elastic cost of defects forced into bundles by rapid quenching~\cite{gov} or by the toroidal topology of bundles form by long strands of DNA~\cite{gelbart}.  In conflict with assumptions of ``electrostatic" mechanisms for limited growth, finite-sized bundles are observed even when bundles are condensed in the absence of multi-valent ions, for example, by depletion forces~\cite{hosek, cebers} or through the incorporation of specialized cross-linking proteins~\cite{purdy, bausch}.  Additionally, finite-diameter bundles of nominally {\it straight} filaments are readily observed, demonstrating that finite-bundle growth cannot as general rule be attributed to the complex topology of the bundle~\cite{kulic_deserno}.  

In view of this phenomenology, it is quite natural to consider the chiral frustration of filament assemblies as a more generic mechanism for limiting bundle growth.  Biological filaments inherit chiral structure from the handedness of the subunits from which they are built, generically imbuing filaments with some measure of helical, screw-like structure.  In this study, we have demonstrated the viability of this mechanism to limit the growth of chiral filament aggregates by way of a very generic model.  Applying the conclusions of the generic continuum-elastic description of hexagonally-packed bundles to a system of particular interest--that is, for biological filaments--requires a detailed accounting of the microscopic physics underlying the interactions between densely-packed filaments as well as the mechanical forces needed to distort them.  Certainly, some information about the phenomenological costs of the various geometrical distortions described in Sec. \ref{sec: elasticity} can be inferred from single-molecule mechanical measurements, or at the least from dimensional arguments.  In particular, recent mechanical measurements of the elastic cost of bending (or unbending) a sinfle {\it salmonella} flagellum allows us to estimate the range of cohesive forces for which bundles should be thermodynamically stable.  The bending stiffness of these flagella is measure to be $C_\kappa = 3.5 {\rm pN \ \mu m^2}$, while the preferred curvature is in the range of $\kappa_0 \simeq 1 {\rm \mu m}^{-1}$.  From eq. (\ref{eq: depc}) we may estimate that bundles of these flagella are stable if the cohesive free energy per unit length is less than $\Delta \epsilon_c \approx k_B T / {\rm nm}$.  

The case of straight bundles of twisted filaments is more directly relevant to the bundle formation of filamentous proteins, such as f-actin or collagen.  Here, it is necessary to determine the values of $\gamma_{{\rm Tw}}$, a measure of the local preference for filament twist.  Computing or measuring the strength of chiral forces in liquid crystalline is a notoriously difficult affair~\cite{harris_kamien_lubensky}, largely due to the effect of positional and rotational fluctuations which considerably weaken the mutual torques exerted by chiral molecules.  The two-dimensional order of bundle geometry simplifies this problem, as the in-plane positional as well as the orientational fluctuations of the filaments are restrained.  The most detailed theory of the strength of chiral forces for biological filaments has been developed by Kornyshev, Leiken and coworkers~\cite{kornyshev_leikin_prl_00, kornyshev_leikin_rmp_07}.  This model treats biological filaments, like DNA, cylindrical rods, along which a helical charge pattern is distributed.  In principle, this model could be adapted to compute the explicit dependence of the preference for a locally-braiding geometry of hexagonally-packed filaments on charge and structure of the helical geometry.  However, it is unlikely the electrostatic forces alone determine the value of $\gamma_{{\rm Tw}}$, since bundled protein filaments are brought to very close separations over order their molecular diameters and smaller.  At this close range, interactions between biological filaments are likely very complex, including contributions from steric and hydrophobic forces between very heterogeneous molecules.  Indeed, atomic force microscopy observations of pairs of intertwined actin filaments~\cite{ikawa_prl_07} suggest that preference for twist for molecules in close contact is quite strong, as filaments readily wind helical around another on length scales much shorter than the molecular persistence length, exposing a state of large bending stress which must be compensated by a lowering of the inter-filmament potential.  Clearly, the development of a predictive and realistic model for the strength of twist-inducing forces in biological filaments is an outstanding problem with significant implications for the self-assembly properties of biological filaments.

\begin{acknowledgments}

I am particularly indebted to R. Bruinsma, with whom some of this work originated.  I would also like to acknowledge R. Kamien, Z. Dogic, T. Powers and T. Witten for many helpful discussions.  This work was supported by UMass, Amherst through a Healey Endowment Grant.  I would also like to acknowledge the hospitality of the Aspen Center for Physics, where some of this work was done.
\end{acknowledgments}
\appendix*

\section{Geometric Decomposition of Chiral Elastic Energy}

Here we demonstrate that the chiral elastic terms allowed by symmetry can be decomposed into two rotationally-invarient measures of the bundle geometry, the {\it twist}, or Tw,  and the {\it writhe}, Wr, of the bundle~\cite{white_69, fuller_pnas_71}.  For a bundle with filaments fixed at the ends or as closed bundle these two quantities are topologically constrained through the theorem,
\begin{equation}
{\rm Lk} = {\rm Tw} + {\rm Wr} , 
\end{equation}
where Lk is the {\it linking number} which counts the number of links between the curves traced out by the bundle backbone and filaments in the bundle.  In the present problem, there are no topological constraints that fix Lk, instead Tw and Wr appear {\it independently} as rotationally-invariant, chiral-symmetry breaking terms.

Consider a filament in the bundle described by the curve, $\rv(s)$, and the $\alpha$th neighboring filament in the hexagonal array that is described by the curve $\rv(s)+\rv_\alpha(s)$.  The twist measures the local rotation of the of $\hat{\rv}_\alpha$ relative to $\rv(s)$.  The integrated twist of these two curves is given by,
\begin{equation}
{\rm Tw}_ \alpha = \frac{1}{2 \pi} \int ds \frac{\tv \cdot(\rv_ \alpha \times \partial_s \rv_ \alpha)}{|\rv_ \alpha |^2} ,
\end{equation}
where $\tv = \partial_s \rv$ is the filament tangent.  Note that $\partial_s \rv_ \alpha =\tv_ \alpha -\tv$  where $\tv_\alpha$ is the tangent to the neighboring filament, which for smoothly-varying filament orientations can be approximated by $\tv_ \alpha \simeq \tv + (\rv_ \alpha \cdot \grad) \tv$.  Using this we compute the average twist by summing over the six neighbors in the hexagonal array,
\begin{equation}
\langle {\rm Tw }\rangle= \frac{1}{12 \pi} \int ds \sum_{\alpha} \tv \cdot  \big[ \hat{\rv}_\alpha \times (\hat{\rv}_\alpha \cdot \grad) \tv \big] ,
\end{equation}
where we have divided by 6 to account for the number of filament pairs per hexagonal plaquette.  Using the result that for a hexagonal lattice $\sum_ \alpha (\hat{r}_\alpha)_i (\hat{r}_\alpha)_j = 3 \delta_{ij}$ (where $i$ and $j$ specify directions in the plane perpendicular to $\tv$) we find the result that,
\begin{equation}
2 \langle {\rm Tw} \rangle =\frac{1}{2 \pi} \int ds ~ \tv\cdot (\grad \times \tv) ,
\end{equation}
which is precisely the nematic twist of the director field.  To lowest order in the displacement of filaments perpendicular to $\hat{z}$ this operator is given by,
\begin{equation}
\tv\cdot (\grad \times \tv) \simeq \hat{z} \cdot (\grad \times \partial_z \uv_\perp) .
\end{equation}
It is straightforward to identify the chiral-energy coupling to twist by expanding terms in eq. (\ref{eq: H*}) to first order in gradients of $\uv_\perp$ to show that the elastic preference to twist is given by the sum of the three chiral elastic parameters, $\gamma_{{\rm Tw}} = \gamma + \gamma' +\gamma''$. 

The writhe of a curve is strictly a {\it global} measure of geometry~\cite{kamien_rmp_02}, requiring a description of the curve at distant points along the backbone.  Using a theorem by Fuller~\cite{fuller_pnas_78}, the writhe may be written as a local quantity by defining it relative to a reference curve.  In this case, it is natural to choose the writhe-free $\hat{z}$ axis as a reference curve, as this represents the orientation of the bundle filaments in the undeformed reference state.  In this case, the writhe of a filament may be written as an integral over a {\it local} quantity as,
\begin{equation}
\label{eq: wr1}
{\rm Wr} = \frac{1}{2 \pi} \int ds \frac{\hat{z} \cdot \big[\tv \times (\tv \cdot \grad) \tv\big]}{ 1 + \tv \cdot \hat{z}} .
\end{equation}
This formula holds provided that $\tv \cdot \hat{z} \neq -1$ everywhere along the contour of the curve.  Again, by expanding the local writhe operator to lowest order in $\uv_\perp$,
\begin{equation}
\frac{\hat{z} \cdot (\tv \times \partial_s \tv)}{ 1 + \tv \cdot \hat{z}} \simeq \frac{1}{2} \hat{z} \cdot (\partial_z \uv_\perp \times \partial^2_z \uv_\perp) ,
\end{equation}
 so that we may associate the higher order contribution from chiral elastic term $\tv \cdot \big[  (\tv \cdot \grad)  \grad \times \uv_\perp \big]$ [from the third term in eq. (\ref{eq: H*})] with filament writhe. From this expansion we deduce that the elastic preference for writhe is simply,  $\gamma_{\rm Wr} = -\gamma''$.  Note that relative to the chiral preference for twist, the elastic coupling to writhe represents a higher-order term in $\uv_\perp$ and derivatives of $\uv_\perp$, which we may regard as symptom of the ``non-local" nature of writhe.   
 


\begin{thebibliography}{99}

\bibitem{harris_kamien_lubensky}
A. B. Harris, R. D. Kamien and T. C. Lubensky, Rev. Mod. Phys. {\bf 71}, 1745 (1999).

\bibitem{crick}
F. H. C. Crick, Nature {\bf 170}, 882 (1952).

\bibitem{wright_mermin}
D. C. Wright and N. D. Mermin, Rev. Mod. Phys. {\bf 61}, 385 (1989).

\bibitem{degennes_ssc_72}
P.-G. de Gennes, Solid State Commun. {\bf 10}, 753 (1972).

\bibitem{renn_lubensky}
S. R. Renn and T. C. Lubensky, Phys. Rev. A {\bf 38}, 2132 (1988).

\bibitem{livolant_leforestier}
F. Livolant and A. Leforestier, Prog. Polym. Sci. {\bf 21}, 1115 (1996).

\bibitem{kamien_nelson_95}
R. D. Kamien and D. R. Nelson, Phys. Rev. Lett. {\bf 74}, 2499 (1995)

\bibitem{kamien_nelson_96}
R. D. Kamien and D. R. Nelson, Phys. Rev. E {\bf 53}, 650 (1996).

\bibitem{prost_helfrich}
W. Helfrich and J. Prost, Phys. Rev. A {\bf 38}, 3065 (1988).

\bibitem{selinger}
J. V. Selinger, M. S. Spector and J. M. Schnur, J. Phys. Chem. B {\bf 105}, 7157 (2001).

\bibitem{ghafouri_bruinsma}
R. Ghafouri and R. Bruinsma, Phys. Rev. Lett. {\bf 94}, 138101 (2005).


\bibitem{grason_bruinsma_07}
G. M. Grason and R. F. Bruinsma, Phys. Rev. Lett. {\bf 99}, 098101 (2007).


\bibitem{degennes_prost}
P. G. de Gennes and J. Prost, {\it The Physics of Liquid Crystals} (Claredon, Oxford, 1993), 2nd ed.

\bibitem{fibrin}
J. W. Weisel, C. Nagaswami and L. Makowski, Proc. Natl. Acad. Sci. USA {\bf 84}, 8991 (1987).

\bibitem{kleman_85}
M. Kl\'eman, J. Physique {\bf 46}, 1193 (1985).

\bibitem{namba}
K. Namba and R. Vonderviszt, Q. Rev. Biophys. {\bf 30}, 1 (1997).

\bibitem{darton_berg_07}
N. C. Darton and H. C. Berg, Biophys. J. {\bf 92}, 2230 (2007).

\bibitem{dogic_fraden}
Z. Dogic and S. Fraden, Cur. Opin. Col. Int. Sci. {\bf 11}, 47 (2006).


\bibitem{tomar_green_day_jacs_07}
S. Tomar, M. M. Green and L. A. Day, J. Am. Chem. Soc. {\bf 129}, 3367 (2007).

\bibitem{barry}
E. Barry, Z. Hensel, Z. Dogic, M. Shribak and R. Oldenbourg, Phys. Rev. Lett. {\bf 96}, 018305 (2006).


\bibitem{starostin}
E. L. Starostin, J. Phys. Condens. Matter {\bf 18}, S187 (2006).

\bibitem{bruinsma_selinger}
J. V. Selinger and R. F. Bruinsma, Phys. Rev. A {\bf 43}, 2910 (1991).


\bibitem{grason_08}
G. M. Grason, Phys. Rev. Lett. {\bf 101}, 105702 (2008).


\bibitem{grelet_prl_08}
E. Grelet, Phys. Rev. Lett. {\bf 100}, 168301 (2008).


\bibitem{landau_lifshitz}
L. D. Landau and E. M. Lifshitz, {\it Theory of Elasticity}, 3rd ed. (Pergamon, Oxford, 1986), Chap. 2.

\bibitem{kamien_jphys}
R. D. Kamien, J. Phys. II France {\bf 6}, 461 (1996).

\bibitem{fuller_pnas_71}
F. B. Fuller, Proc. Nat. Acad. Sci. U.S.A. {\bf 68}, 815 (1971). 


\bibitem{liang_upmanyu}
H. Y. Liang and M. Upmanyu, Carbon {\bf 43}, 3189 (2005).

\bibitem{kamien_rmp_02}
R. D. Kamien, Rev. Mod. Phys. {\bf 74}, 953 (2002).

\bibitem{israelachvili}
J. Israelachvili, {\it Intermolecular and Surface Forces}, (Academic Press, London, 2006), Chap. 16.


\bibitem{wong_prl_07}
G. H. Lai, R. Coridan, O. V. Zribi, R. Golestanian and G. C. L. Wong, Phys. Rev. Lett. {\it in press} (2007).


\bibitem{kleman_05}
M.-F. Achard, M. Kl\'eman, Y. A. Nastishin and H.-T. Nguyen, Eur. Phys. J. E {\bf 16}, 37 (2005).


\bibitem{alberts}
B. Alberts, D. Bray, J. Lewis, M. Raff and K. Roberts, {\it Molecular Biology of the Cell}, 4th ed. (Garland Science, New York, 2002).

\bibitem{theriot}
S. M. Rafelski and J. A. Theriot, Annu. Rev. Biochem. {\bf 73}, 209 (2004).


\bibitem{tang_pchem_96}
J. X. Tang, S. Wong, P. T. Tran and P. A. Janmey, Ber. Bunsenges. Phys. Chem. {\bf 100}, 796 (1996).


\bibitem{pelletier_prl_03}
O. Pelletier, E. Pokidysheva, L. S. Hirst, N. Bouxsein, Y. Lin and C. R. Safinya, Phys. Rev. Lett. {\bf 91}, 148192 (2003).

\bibitem{hosek}
M. Hosek and J. X. Tang, Phys. Rev. E {\bf 69}, 051907 (2004).

\bibitem{cebers}
A. Cebers, Z. Dogic and P. A. Janmey, Phys. Rev. Lett. {\bf 96}, 247801 (2006).

\bibitem{purdy}
K. R. Purdy, J. R. Bartles and G. C. L. Wong, Phys. Rev. Lett. {\bf 98}, 058105 (2007).

\bibitem{bausch}
M. M. A. E. Claessens, C. Semmrich, L. Ramos and A. R. Bausch, Proc. Natl. Acad. Sci. USA {\bf105}, 8819 (2008). 

\bibitem{ha_liu}
B.-Y. Ha and A. J. Liu, Europhys. Lett. {\bf 46}, 624 (1999).


\bibitem{henle}
M. Henle and P. A. Pincus, Phys. Rev. E {\bf 71}, 060801(R) (2005).

\bibitem{gov}
N. S. Gov, Phys. Rev E. {\bf 78}, 011916 (2008).

\bibitem{gelbart}
S. Y. Park, D. Harries and W. M. Gelbart, Biophys. J. {\bf 75}, 714 (1998).

\bibitem{kulic_deserno}
I. M. Kuli\'c, D. Andienko and M. Deserno, Europhys. Lett. {\bf 67}, 418 (2004).




\bibitem{kornyshev_leikin_prl_00}
A. A. Kornyshev and S. Leikin, Phys. Rev. Lett. {\bf 84}, 2537 (2000).

\bibitem{kornyshev_leikin_rmp_07}
A. A. Kornyshev, D. J. Lee, S. Leikin and A. Wynveen, Rev. Mod. Phys. {\bf 79}, 943 (2007).

\bibitem{ikawa_prl_07}
T. Ikawa, F. Hoshino, O. Watanabe, Y. Li, P. Pincus and C. R. Safinya, Phys. Rev. Lett. {\bf 98}, 018101 (2007).




\bibitem{fuller_pnas_78}
F. B. Fuller, Proc. Nat. Acad. Sci. U.S.A. {\bf 75}, 3557 (1978). 

\bibitem{white_69}
J. H. White, Am. J. Math. {\bf 91}, 693 (1969).

\end{thebibliography}
\end{document}